\documentstyle[sprocl,epsfig]{article}

\def\be{\begin{equation}}
\def\ee{\end{equation}}
\def\bea{\begin{eqnarray}}
\def\eea{\end{eqnarray}}

\newcommand{\AmS}{{\protect\the\textfont2
  A\kern-.1667em\lower.5ex\hbox{M}\kern-.125emS}}
\newcommand{\lsim}{\raisebox{-3pt}{$\,\stackrel{\textstyle <}{\sim}\,$}}
\newcommand{\gsim}{\raisebox{-3pt}{$\,\stackrel{\textstyle >}{\sim}\,$}}

\def\br{{\bf r}}
\def\brp{{\bf r^{\prime}}}

\def\Pom{{\bf I\!P}}

\def\lsim{\mathrel{\rlap{\lower4pt\hbox{\hskip1pt$\sim$}}
    \raise1pt\hbox{$<$}}}         

\def\gsim{\mathrel{\rlap{\lower4pt\hbox{\hskip1pt$\sim$}}
    \raise1pt\hbox{$>$}}}         

\begin{document}

\title{
Small-x Structure Functions and QCD Pomeron}



\author{\underline{N.N. Nikolaev}$^{a,b)}$ \& V.R. Zoller$^{c}$\bigskip\\}

\address{$^{a)}$Institut f. Kernphysik, Forschungszentrum J\"ulich,
D-52450 J\"ulich, Germany\\
$^{b)}$ L.D.Landau Institute for Theoretical Physics,
142432 Chernogolovka, Russia\\
$^{c)}$ Institute for Theoretical and Experimental Physics, Moscow, Russia} 


\maketitle\abstracts{ 
The recent progress in color BFKL-Regge phenomenology of
small-$x$ DIS on nucleons, pions and photons is reviewed}

 As noticed by Fadin, Kuraev and Lipatov \cite{FKL}
 and discussed in detail by Lipatov \cite{Lipatov},
the incorporation of asymptotic freedom (AF), i.e.  the
 running QCD coupling, into the BFKL equation makes 
the QCD pomeron a series of moving Regge poles.
The trajectories if the running BFKL poles have been
calculated by us and B.G.Zakharov in the color dipole (CD)
approach to running BFKL equation \cite{JETPLett,DER}.
 The contribution of the each pole to scattering
 amplitudes satisfies the seasoned Regge factorization
which is a basis of our BFKL-Regge expansion for small-$x$
DIS in the CD basis  \cite{JETPLett,DER}. We review 
here recent applications of the BFKL-Regge factorization to DIS off
pions and photons and the charm structure function of the proton.
More details and references to early works on the subject are 
found in \cite{NSZpion,NSZgamgam,NZcharm}.

In the CD basis the beam-target interaction is viewed as a scattering 
of color dipoles $\bf r$ and
 $\bf{r^{\prime}}$ in both  the beam ($b$) and target ($t$) particles.
Once the beam and target independent dipole-dipole cross section
$\sigma(x,\bf{r},\bf{r^{\prime}})$
is known one can calculate
 $\sigma^{bt}(x)$
 making use of the
 CD factorization
\be
\sigma^{bt}(x)
=\int dz d^{2}{\bf{r}} dz^{\prime} d^{2}{\bf{r^{\prime}}}
|\Psi_b(z,{\bf{r}})|^{2} |\Psi_{t}(z^{\prime},{\bf{r^{\prime}}})|^{2}
\sigma(x,{\bf{r},\bf{r^{\prime}}}) \,,
\label{eq:1.3}
\ee
where $|\Psi_b(z,{\bf{r}})|^{2}$ and
$|\Psi_{t}(z^{\prime},{\bf{r^{\prime}}})|^{2}$
 are probabilities to find
 a CD,
$\br$ and $\brp$ in the beam and target, respectively.
Here we emphasize that all the beam and target dependence is
contained
in the CD distributions $|\Psi_b(z,\br)|^2$ and $|\Psi_t(z,\br)|^2$.
 The CD BFKL-Regge factorization
uniquely prescribes\cite{DER} an expansion of the
 dipole-dipole total cross section
$\sigma(x,{r},{r^{\prime}})$ in eigen-functions $\sigma_m(r)$ of 
the CD BFKL equation
\be
\sigma(x,r,r^{\prime})=\sum_{m}C_m\sigma_m(r)\sigma_m(r^{\prime})
\left({x_0\over x}\right)^{\Delta_m}\,.
\label{eq:PRODUCT}
\ee

The AF exacerbates the well known infrared sensitivity
of the CD BFKL equation and infrared regularization is called upon:
infrared freezing of $\alpha_S$ and finite propagation radius $R_c$
of perturbative gluons were consistently used in our CD
approach to BFKL equation since 1994 \cite{NZDelta}. The past years the both concepts
 have become widely accepted.

The leading eigen-function $\sigma_0(r)$ for the ground state with 
intercept $\Delta_0\equiv\Delta_{\Pom}$
 is node free, subleading $\sigma_m(r)$ have 
 $m$ radial  nodes.  With our infrared regulator
the intercept of the leading pole
  trajectory is  found to be  $\Delta_{\Pom}=0.4$ and $\Delta_m$ are
found to follow closely the Lipatov's law\cite{Lipatov}
$ \Delta_m= {\Delta_0/(m+1)}$. For our infrared regulator
 the first node of $\sigma_{1}(r)$ is located at
$r=r_1\simeq 0.05-0.1\,{\rm fm}$,  for solutions with
$m\geq 4$ the  higher nodes are located at a very small $r$  
way beyond the resolution
scale
$1/\sqrt{Q^2}$. In practical evaluation of $\sigma^{bt}$
 we can truncate expansion (2) at $m=3$ lumping in the term
with $m=3$ contributions of all singularities with $m\geq 3$.

The expansion coefficients $C_m$ in eq.(2) are fully determined
by the boundary condition $\sigma(x_0,r,r^{\prime} )$.
The very ambitious program of description of $F_{2}^{p}(x,Q^2)$
starting
 from  the, perhaps  excessively restrictive, but appealingly natural,
 two-gluon exchange boundary condition at $x_0=0.03$ has been launched by us in
\cite{NZDelta} and met  with  remarkable phenomenological success
\cite{DER,NSZpion}.
\begin{figure}[!htb]
\epsfig{file=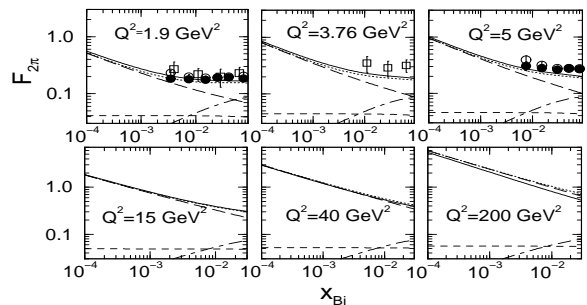, height = 1.6in, width = 3in}
\vspace{-0.5cm}
\caption{
Predictions  from the CD BFKL-Regge factorization for the pion
 structure function
$F_{2\pi}(x_{Bj},Q^2)$ (solid lines) vs.
the experimental data from the H1 \protect\cite{F2PI}  are shown.
The different components of $F_{2\pi}(x_{Bj},Q^2)$ are shown:
 the valence contribution (dashed-dotted),
 the non-perturbative soft contribution (dashed), 
the Leading Hard+Soft+Valence Approximation
 $F^{\rm LHSVA}_{2\pi}(x_{Bj},Q^2)$ (dotted) and the sum of the soft and
 leading hard pole contributions without valence component (long dashed).}
\label{fig:1}
\end{figure}

The exchange by perturbative gluons is a dominant mechanism for small dipoles
$r\lsim R_c$, interaction of large dipoles
is modeled by the non-perturbative, soft mechanism which we
approximate here by a factorizable soft pomeron with intercept
$\alpha_{\rm soft}(0)-1=\Delta_{\rm soft}=0$, i.e., flat vs. $x$ at
small $x$. Then the extra term
$C_{\rm soft}\sigma_{\rm soft}(r)\sigma_{\rm soft}(r^{\prime})$
must be added in the  r.h.s. of expansion (\ref{eq:PRODUCT}). 
At moderately small $x$ we
include a contribution from DIS on valence  quarks.

\begin{figure}[!htb]
\epsfysize 1.5 in 
\epsfbox{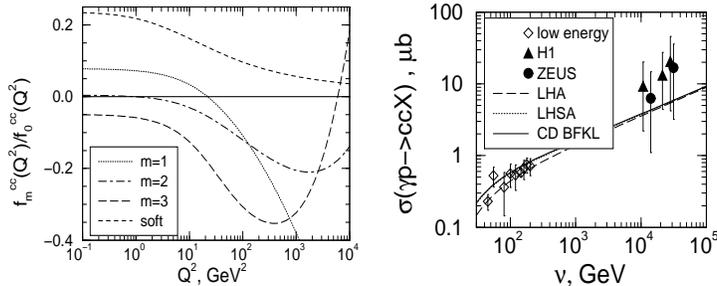}
\vspace{-0.5cm}
\caption{{\bf The left box:} The subleading hard-to-rightmost 
hard and soft-pomeron-to-rightmost hard 
ratio of eigen-structure functions $f_m^{cc}(Q^2)/f_0^{cc}(Q^2)$  as a function 
$Q^{2}$.~~
{\bf The right box:}
Predictions from CD BFKL-Regge factorization for  $\sigma(\gamma p\to c{\bar c}X)$
vs the experimental data compiled in \protect\cite{DATAHERACC}.
The  solid curve is a complete
CD BFKL-Regge expansion, the long-dashed line is the rightmost hard BFKL pole  
(LHA), the dotted curve
ia a sum of the rightmost hard BFKL plus
soft-pomeron exchanges (LHSA).  
 }
\label{fig:2}
\end{figure}

It is convenient to introduce  the  eigen
structure functions $(m={\rm soft},0,1,2,...)$
\be
f_m(Q^2)= {Q^2\over
 {4\pi^2\alpha_{em}}}
 \sigma_m^{\gamma^*}(Q^2)\, ,
\label{eq:F2CC}
\ee
where $
\sigma_m^{\gamma^*}(Q^2)=
 \langle {\gamma^*}  | \sigma_m(r)|{\gamma^*}$. Then ($m={\rm soft},0,1,2,...$)
\bea
\sigma^{\gamma^* t}(x,Q^2)=\sum_{m} C_m\sigma^{\gamma^*}_m(Q^2)\sigma^{t}_m
\left({x_0\over x}\right)^{\Delta_m}+\sigma^{\gamma^* t}_{\rm val}(x,Q^2)
\,,
\label{eq:PRODUCT1}
\eea
\be
F_{2t}(x,Q^2)=\sum A^{t}_m
f_m(Q^2)
\left({x_0\over {x}}\right)^{\Delta_m}
+F_{2t}^{\rm val}(x,Q^2)\,,
\label{eq:F2TRUN}
\ee
where the target ($t=p,\pi,\gamma,\gamma^*...)$ dependence comes exclusively from
$
\sigma^{t}_{m}=\langle t|\sigma_m(r)|t\rangle=
\int dz d^{2}{\bf{r}}
 |\Psi_{t}(z,r)|^{2}
\sigma_m(r) \,.$

The CD BFKL approach predicts uniquely that for light flavours 
sub-leading eigen structure 
functions  $f_{m\geq 1}(Q^2)$ have their first node at $Q^{2} \sim 20-60$ 
GeV$^{2}$ \cite{JETPLett,DER}. In this range of $Q^2$ the proton, pion and
real photon structure functions are well reproduced by the Leading Hard$+$ 
Soft$+$Valence Approximation (LHSVA)
which gives a unique handle on the intercept $\alpha_{\Pom}(0)=1+\Delta_0$
of the leading hard BFKL pole. This point is illustrated in fig.~1
where the dotted curve represents the LHSVA for the pion structure function
and is hardly distinguishable from the solid curve for the full fledged
BFKL-Regge expansion.

The numerical results for pion can be well approximated by $
F_{2\pi}(x,Q^2)\simeq {2\over 3}F_{2p}\left({2\over 3}x,Q^2\right)$.
This additive quark rule derives for the hard component from $R_{c}^{2} \ll
\langle r_{p}^{2}\rangle,\langle r_{\pi}^{2}\rangle$, whereas for the soft 
component it derives from approximate equality of the quark-quark separation
proton and quark-antiquark separation in the pion. The agreement with the
recent H1 data \cite{F2PI} is very good.

\begin{figure}[!htb]
\epsfig{file=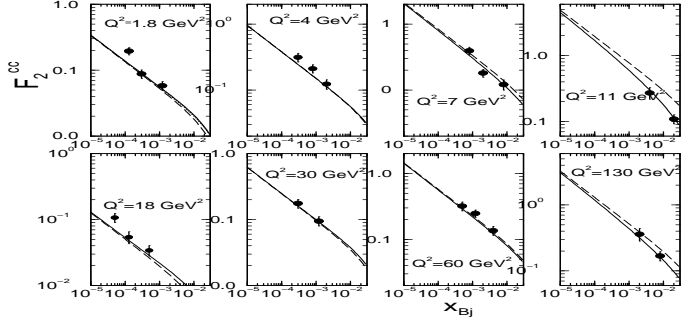, height = 1.7in, width = 3.5 in}
\vspace{-0.4cm}
\caption{
Prediction from CD BFKL-Regge factorization for the  charm structure function 
of the proton $F^{cc}_2(x_{Bj},Q^2)$ vs. the experimental data from 
 ZEUS  \protect\cite{ZEUScc}.  The legend of curves is the same as in fig. 1. }
\label{fig:3}
\end{figure}

In CD approach open charm is excited from $c\bar{c}$ color
dipoles of small size,
\be
{ 4 \over Q^{2}+4m_{q}^{2}}\lsim r^{2} \lsim { 1\over m_{q}^{2}}\, 
\label{eq:1.1}
\ee
which for a broad range of $Q^2 \lsim 100$ GeV$^2$ 
is close to the position of the first
node of subleading $\sigma_m(r)$. 
Because the soft contribution to
charm production is negligible small, this entails the dominance by rightmost
hard BFKL pole, see fig. 2. Here the {\sl l.h.s} box shows the ratio of
subleading-to-rightmost BFKL pole. 
Recall that this point about charm excitation being a clean probe of the hard most
BFKL pole has been made by us and B.G.Zakharov already
in 1994 \cite{NZDelta}. 
 The found nice agreement with the experimental data from ZEUS
Collaboration \cite{ZEUScc} on the charm 
structure function of the proton (fig. 3) and open charm
photoproduction \cite{DATAHERACC} (fig. 2)
strongly corroborates our 1994 prediction 
$\Delta_{\Pom}=\alpha_{\Pom}(0)-1 \approx 0.4$ for the intercept of the 
rightmost hard BFKL pole \cite{NZDelta}. 
\begin{figure}[!htb]
\epsfig{file=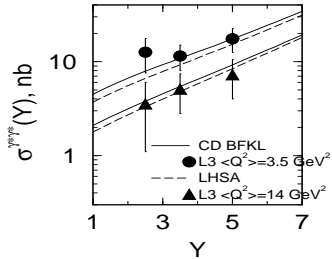, height=1.4in, width = 1.7in}
\vspace{-.3cm}
\caption{
Predictions from CD BFKL-Regge expansion for the vacuum exchange in  
$\gamma^{*}\gamma^{*}$ cross section at $\langle Q^{2} \rangle = 
\langle P^{2}\rangle $  vs. the experimental data from L3 \protect\cite{L3SIGY}. 
The variable $Y=\log(W^2/\sqrt{Q^2P^2})$.
}
\label{fig:5}
\end{figure}

The virtuality $P^{2}$ of the `target' photon in $\gamma^*(Q^2)\gamma^*(P^2)$ 
scattering offers still further tests of BFKL-Regge factorization, which
gives the parameter free prediction ($m={\rm soft},0,1,2,...$)

\bea
\sigma^{\gamma^*\gamma^*}(x,Q^2,P^2)= \hfill\nonumber \\
{(4\pi^2 \alpha_{em})^{2} \over Q^{2}P^{2}}
\sum_{m} C_m f_m(Q^2)f_m(P^2)
\left({3x_0\over 2x}\right)^{\Delta_m}
+\sigma^{\gamma^*\gamma^* }_{\rm qval}(x,Q^2,P^2)\,.
\label{eq:2.5}
\eea
The quasi-valence (reggeon) correction is important
at not so small $x$.
\begin{figure}[!htb]
\epsfig{file=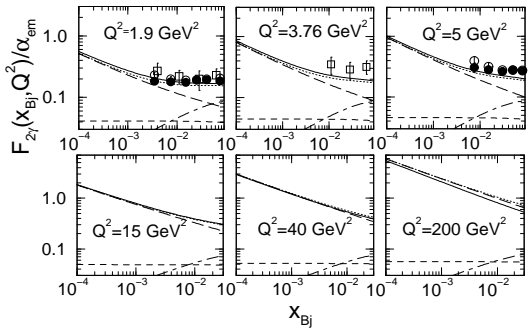, height=1.8in, width = 2.7 in}
\vspace{-.3cm}
\caption{
Predictions from CD BFKL-Regge expansion  for the photon structure function.
 The legend of curves is the same as in fig.1. The data 
points are from \protect\cite{L3,OPAL} 
 }
\label{fig:4}
\end{figure}

The CD BFKL approach predicts  
uniquely that that because of the node effect
at $Q^{2}\sim 20$ GeV$^2$ in which region of $Q^{2}$ the rightmost
hard BFKL pole contribution will dominate. 
Recently the L3 collaboration \cite{L3SIGY} reported the first experimental 
evaluation of the vacuum exchange in equal virtuality $\gamma^{*}
\gamma^{*}$ scattering. Their procedure of subtraction of the 
non-vacuum reggeon and/or the Quark Parton Model contribution
is described in \cite{L3SIGY}, arguably 
the subtraction
uncertainties are marginal within the present error bars. In fig.~4 we
compare our predictions to the L3 data \cite{L3SIGY} shown vs. 
the variable $Y=\log(W^2/\sqrt{Q^2P^2})$.
The agreement of our estimates 
with the experiment is good, the contribution of subleading hard BFKL 
exchange is negligible within the experimental error bars.

Our predictions for the photon structure function are parameter-free 
and are presented in 
fig.~5 in comparison with the recent  L3 and OPAL  data \cite{L3,OPAL}.
 A comparison of the solid and 
dotted curves shows clearly that subleading hard BFKL exchanges are
numerically small in the experimentally interesting region of $Q^{2}$,
the rightmost hard BFKL pole exhausts the hard vacuum contribution 
for $2 \lsim  Q^{2} \lsim 100\, $ GeV$^{2}$. \\

\noindent
{\bf Conclusions:} The BFKL-Regge factorization has lead to a remarkable progress
in relating parameter-free  to each other structure functions of various targets.
The understanding of nodal properties of eigenfunctions of CD BFKL equation
has shed a light on when and why the hard contribution to structure functions
is dominated by the contribution from the rightmost hard BFKL pole. 
There is a mounting evidence for the intercept of the rightmost
hard BFKL pole $\Delta_{\Pom}=0.4$ as predicted by us in 1994. \\ 

\noindent
{\bf Acknowledgments: } I'm indebted to Herb Fried and Chung-I Tan 
for invitation to QCD2000. This work was partly supported by the grants
INTAS-96-597 and INTAS-97-30494 and DFG 436RUS17/11/99.

\end{document}